\documentclass{article}
\usepackage[utf8]{inputenc}
\usepackage{amsthm}
\usepackage{graphicx,ae,aecompl}
\usepackage{amssymb}
\usepackage{amsmath}
\usepackage{subfigure}
\usepackage{algorithmic}
\usepackage[boxruled,linesnumbered]{algorithm2e}
\usepackage{makeidx}
\usepackage{float}
\usepackage{scalefnt}
\usepackage{indentfirst}

\usepackage[left=2.5cm,top=2.5cm,right=2.5cm,bottom=2.5cm]{geometry}

\title{A bayesian wavelet shrinkage rule under LINEX loss function}
\author{Alex Rodrigo dos S. Sousa \\ State University of Campinas, Brazil \\ asousa@unicamp.br}
\date{}

\begin{document}

\numberwithin{equation}{section}
\numberwithin{table}{section}
\numberwithin{figure}{section}

 \maketitle
    \begin{abstract}
This work proposes a wavelet shrinkage rule under asymmetric LINEX loss function and a mixture of a point mass function at zero and the logistic distribution as prior distribution to the wavelet coefficients in a nonparametric regression model with gaussian error. Underestimation of a significant wavelet coefficient can lead to a bad detection of features of the unknown function such as peaks, discontinuities and oscillations. It can also occur under asymmetrically distributed wavelet coefficients. Thus the proposed rule is suitable when overestimation and underestimation have asymmetric losses. Statistical properties of the rule such as squared bias, variance, frequentist and bayesian risks are obtained. Simulation studies are conducted to evaluate the performance of the rule against standard methods and an application in a real dataset involving infrared spectra is provided.    
        
    \end{abstract}

\section{Introduction}
Wavelet shrinkage methods are important tools to be applied in nonparametric regression models under wavelet basis expansion of the unknown function. In the wavelet domain, coefficients of the representation of a function are typically sparse and the few nonzero coefficients are localized at important positions of the function to be recovered, such as peaks, discontinuities and oscillations. However the observed wavelet coefficients, usually called empirical coefficients which are obtained by the application of a discrete wavelet transformation on the data, are not sparse in practice due the presence of noise. In this sense, wavelet shrinkage rules act by denoising the empirical coefficients and estimating the true wavelet coefficients of the representation of the unknown function. The concept of wavelet shrinkage was introduced in the seminal works by Donoho (1993, 1995), Donoho and Johnstone (1994, 1995) and Donoho et al. (1995).

There are several wavelet shrinkage rules available in the literature, most of them are thresholding. A thresholding rule shrinks a small empirical coefficient that is less than a threshold value to exactly zero. The hard and soft thresholding rules of Donoho and Johnstone (1994) are the most famous and applied in the literature and policies to obtain the threshold value were proposed since then, such as the universal threshold by Donoho and Johnstone (1994), the Stein unbiased risk estimator of Donoho and Johnstone (1995) and the cross validation threshold by Nason (1996), among others. Bayesian wavelet shrinkage rules have also been proposed along the last decades. In fact, bayesian procedures allow the incorporation of prior information about the wavelet coefficients in the estimation process such as sparseness level and boundedness (if they are limited) through the elicitation of a prior probabilistic distribution to them. Chipman et al. (1997) proposed a mixture of two zero mean normal distributions as prior, one of them with very small variance, in the same idea of a spike and slab prior. Mixtures of a point mass function at zero and a symmetric and unimodal distribution centered at zero were proposed by Vidakovic and Ruggeri (2001) with the Laplace distribution, the Bickel distribution by Angelini and Vidakovic (2004), the double Weibull distribution by Reményi and Vidakovic (2015), the logistic distribution by Sousa (2020) and the beta distribution by Sousa et al. (2020).

Although the bayesian shrinkage rules mentioned above and others are well succeeded in many real data applications, they were built under symmetric loss functions, in particular, the quadratic loss function, which implies that overestimation and underestimation with the same magnitudes of a given wavelet coefficient have the same loss. Moreover, when the quadratic loss function is considered, mathematical advantages are gained such as differentiation and easier analytical manipulation, in addition to the fact that the Bayes rule is the posterior expected mean of the wavelet coefficient. However, when overestimation and underestimation should have different losses, the proposed shrinkage rules are not suitable and it is necessary the development of a shrinkage rule that is built under an asymmetric loss function. In this sense, this paper proposes a bayesian shrinkage rule under a point mass function at zero and the logistic distribution as prior to the wavelet coefficient and that is obtained under the so called linear-exponential (LINEX) loss function. This asymmetric loss function that was originally proposed by Varian (1975) penalizes exponentially the underestimation (overestimation) and linearly the overestimation (underestimation) according to convenient choices of its parameters. Under the wavelet estimation point of view, its application is interesting once underestimation of significant wavelet coefficients could not detect relevant features of the unknown function, such as peaks and/or discontinuities for example. Huang (2012) and Torehzadeh and Arashi (2014) proposed policies to obtain the threshold value to the soft thresholding rule under LINEX loss function with a bayesian interpretation of the estimation process, however the bayesian shrinkage rule to be proposed in this work allows prior information about the wavelet coefficients to be taken into account in the bayesian inference about them.   

This paper is organized as follows: the statistical model is defined in Section 2. The estimation procedure, including the proposed shrinkkage rule under LINEX loss, is described in Section 3. Section 4 is dedicated to simulation studies that were conducted to evaluate the performance of the proposed rule. An application in a real dataset about infrared spectra denoising is done in Section 5. Final considerations are in Section 6.  

\section{Statistical Model}

We start with the unidimensional nonparametric regression problem, i.e, one observes $n = 2^J$ points $(x_1,y_1),\cdots, (x_n,y_n)$, $J \in \mathbb{N}$, from the model
\begin{equation}\label{dmodel}
y_i = f(x_i) + e_i,
\end{equation}
where $x_1, \cdots, x_n$ are equidistant scalars, $f \in \mathbb{L}^2(\mathbb{R}) = \{f:\int f^2 < \infty \}$ is an unknown function and $e_1, \cdots, e_n$ are independent zero mean normal random errors with unknown common variance $\sigma^2$, $\sigma > 0$. The goal is to estimate the function $f$ from the sample $\{(x_i,y_i)\}_{i=1}^{n}$ without any assumption regarding the functional structure of $f$, except that it is squared integrable. 

\begin{figure}[H]
\centering
\includegraphics[scale=0.70]{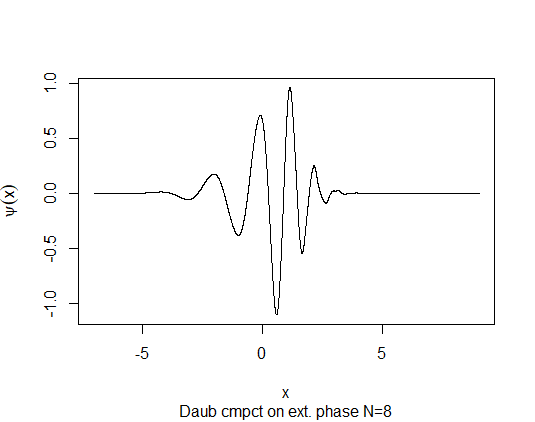}
\caption{Daubechies' compactly supported orthogonal wavelet with eight null moments.}\label{fig:wav}
\end{figure}

The classical approach to estimate $f$ under the nonparametric point of view is to expand it in terms of some functional basis, such as polynomials, Fourier, splines and wavelets for example, see Takezawa (2005). In this work we represent $f$ in model \eqref{dmodel} in terms of wavelet basis, 
\begin{equation} \label{expan}
f(x) = \sum_{j,k \in \mathbb{Z}}\theta_{j,k} \psi_{j,k}(x), 
\end{equation}
where $\{\psi_{j,k}(x) = 2^{j/2} \psi(2^j x - k),j,k \in \mathbb{Z} \}$ is an orthonormal wavelet basis for $\mathbb{L}_2(\mathbb{R})$ constructed by dilations $j$ and translations $k$ of a function $\psi$ called wavelet or mother wavelet and $\theta_{j,k}$ are wavelet coefficients that describe features of $f$ at spatial location $2^{-j}k$ and scale $2^j$ or resolution level $j$. Note that the problem of estimating $f$ becomes a problem of estimating the wavelet coefficients in \eqref{expan}. In fact, wavelet is a function that satisfies the admissibility condition $\int |\Psi(\omega)|^2/|\omega| < \infty$, where $\Psi(\cdot)$ is the Fourier transformation of $\psi$. The admissibility condition implies that $\int \psi = 0$, which motivates the name wavelet. There are several wavelet functions available in the literature, see Vidakovic (1999). For the development of this work, we considered the Daubechies' compactly supported orthogonal wavelet, that does not have closed form but it is shown in Figure \ref{fig:wav}.
We can rewrite the model \eqref{dmodel} in vector notation,
\begin{equation}\label{vmodel}
\boldsymbol{y} = \boldsymbol{f} + \boldsymbol{e},
\end{equation} 
where $\boldsymbol{y} = [y_1,\cdots,y_n]'$, $\boldsymbol{f} = [f(x_1),\cdots,f(x_n)]'$ and $\boldsymbol{e} = [e_1,\cdots,e_n]'$. We apply a wavelet discrete transform (DWT) on the original data to take them to the wavelet domain. Although DWT is usually performed by fast algorithms such as the pyramidal algorithm, it is possible to represent it by a transformation matrix $\boldsymbol{W}$ with dimension $n \times n$ which is applied on both sides of \eqref{vmodel}. Once DWT is linear, we obtain the following model in wavelet domain
\begin{equation}\label{wmodel}
\boldsymbol{d} = \boldsymbol{\theta} + \boldsymbol{\varepsilon},
\end{equation}
where $\boldsymbol{d} = \boldsymbol{W}\boldsymbol{y} = [d_1, \cdots, d_n]'$ is the vector of empirical (observed) wavelet coefficients, $\boldsymbol{\theta} = \boldsymbol{W}\boldsymbol{f} = [\theta_1, \cdots, \theta_n]'$ is the sparse vector of unknown wavelet coefficients of $f$ and $\boldsymbol{\varepsilon} = \boldsymbol{W}\boldsymbol{e} = [\varepsilon_1,\cdots,\varepsilon_n]'$ is the vector of random errors. Thus we can see the empirical wavelet coefficients $\boldsymbol{d}$ as noised versions of the unknown wavelet coefficients  $\boldsymbol{\theta}$. Further, the random errors  in the wavelet domain $\varepsilon_1,\cdots,\varepsilon_n$ remain independent zero mean normal with common variance $\sigma^2$ since the DWT is orthogonal. We will drop the subindices and consider the model $d = \theta + \varepsilon$ for a single wavelet coefficient along the text for simplicity.

The estimation of the vector of wavelet coefficients $\boldsymbol{\theta}$ will be done coefficient by coefficient under a bayesian framework. In this sense, a prior distribution is assigned to a single wavelet coefficient $\theta$ that incorporates our knowledge about its sparsity and symmetry around zero. Then, we consider the prior distribution $\pi(\cdot;\alpha,\tau)$ based on a mixture of a point mass function at zero $\delta_0(\cdot)$ and a symmetric around zero logistic distribution $g(\cdot;\tau)$ proposed by Sousa (2020),
\begin{equation}\label{prior}
\pi(\theta;\alpha,\tau) = \alpha \delta_0(\theta) + (1-\alpha)g(\theta;\tau),
\end{equation}   
where
\begin{equation}
g(\theta;\tau) = \frac{\exp\{-\theta/\tau\}}{\tau(1 + \exp\{-\theta/\tau\})^2}\mathbb{I}_{\mathbb{R}}(\theta), \nonumber
\end{equation}
$\alpha \in (0,1)$, $\tau > 0$ and $\mathbb{I}_{\mathbb{R}}(\cdot)$ is the indicator function on the real set $\mathbb{R}$. Thus, the prior distribution has two hyperparameters to be elicited, $\alpha$ and $\tau$. As discussed in Sousa (2020), these hyperparameters control the severity of the shrinkage imposed by the bayesian rule on the empirical coefficients. In fact, the empirical coefficients are shrunk more for
higher values of $\alpha$ and smaller values of $\tau$ since the prior distribution becomes more concentrated around zero. In the same work, the author suggests values of $\tau \leq 10$ and the elicitation of $\alpha$ according to Angelini and Vidakovic (2004),
\begin{equation}\label{eq:alpha}
\alpha = \alpha(j) = 1 - \frac{1}{(j-J_{0}+1)^\gamma},
\end{equation}
where $J_ 0 \leq j \leq J-1$, $J_0$ is the primary resolution level, $J$ is the number of resolution levels, $J=\log_{2}(n)$ and $\gamma > 0$. They also suggest that in the absence of additional information, $\gamma = 2$ can be adopted.
Finally, it is necessary to estimate $\sigma$ for a complete (hyper)parameter elicitation. According to Donoho and Johnstone (1994), based on the fact that much of the noise information present in the data can be obtained on the finer resolution scale, they proposed
\begin{equation}\label{eq:sigma}
\hat{\sigma} = \frac{\mbox{median}\{|d_{J-1,k}|:k=0,...,2^{J-1}\}}{0.6745}.
\end{equation}

\section{Estimation Procedure}
The estimation of the wavelet coefficients vector $\boldsymbol{\theta}$ is done by the application of the bayesian wavelet shrinkage rule $\delta^{*}(\cdot)$ associated to the models \eqref{wmodel} and \eqref{prior} on the empirical wavelet coefficients vector $\boldsymbol{d}$, i.e, for a single wavelet coefficient $\theta$, its estimator $\hat{\theta}$ is 
\begin{equation}
\hat{\theta} = \delta^{*}(d).
\end{equation}
In fact the estimator $\hat{\theta}$ is a shrunk (denoised) version of the empirical coefficient $d$. Once we have the estimated wavelet coefficients vector $\boldsymbol{\hat{\theta}} = [\hat{\theta}_1,\cdots,\hat{\theta}_n]'$, the function values $\boldsymbol{f}$ is then estimated by the inverse discrete wavelet transformation (IDWT) application that can be represented by the transpose matrix $\boldsymbol{W'}$, 
\begin{equation}\label{idwt}
\boldsymbol{\hat{f}} = \boldsymbol{W'}\boldsymbol{\hat{\theta}}.
\end{equation}
The bayesian rule $\delta^{*}(\cdot)$ is obtained by minimizing the posterior expected loss $\mathbb{E}_{\pi}[L(\delta,\theta)|d]$, i.e
\begin{equation}
\delta^{*}(d) = \mathrm{arg} \min_{\delta} \mathbb{E}_{\pi}[L(\delta,\theta)|d],
\end{equation}
according to a loss function specification $L(\delta,\theta)$, which is done depending on the estimation problem and its costs. For symmetric losses as example, the well known quadratic loss function $L(\delta,\theta) = (\delta - \theta)^2$ is generally applied and its associated Bayes rule is the posterior expected mean $\delta^{*}(d) = \mathbb{E}_{\pi}(\theta|d)$, see Robert (2007) for a complete development of bayesian frameworks. In this work however, we consider wavelet coefficients estimation problem under asymmetric losses, in the sense that underestimation and overestimation of $\theta$ lead to different costs. Thus, the LINEX loss function is a very suitable function to be considered in this context. The next subsections describe the LINEX loss function, the associated Bayes rule and its statistical properties.

\subsection{LINEX Loss Function}
We consider the unidimensional LINEX loss function by Varian (1975) given by
\begin{equation}\label{loss}
L(\delta,\theta) = b[e^{a(\delta - \theta)} - a(\delta - \theta) - 1],
\end{equation}
for $a \neq 0$ and $b > 0$. The sign of $a$ indicates the direction of the exponential and linear losses. In fact, for $a > 0$, the loss is exponential when $\delta > \theta$ and linear when $\delta < \theta$, i.e, the loss is greater in overestimation. On the other hand, for $a < 0$, the loss is exponential when $ \delta < \theta$ and linear when $\delta > \theta$, i.e, the loss function penalizes more the underestimation. The parameter $b$ controls the general magnitudes of the losses. 

\begin{figure}[H]
\centering
\subfigure[LINEX functions for $b=1$.]{
\includegraphics[scale=0.55]{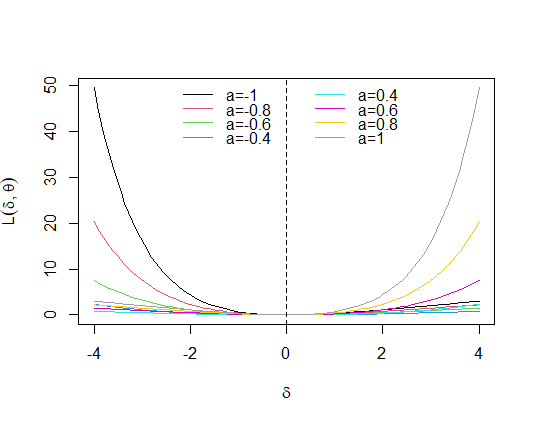}}
\subfigure[LINEX functions for $a=1$.]{
\includegraphics[scale=0.55]{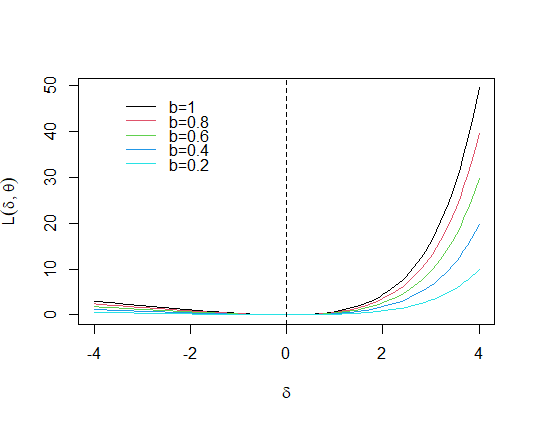}}
\caption{LINEX loss functions $L(\delta,0)$ (i.e the loss when $\theta = 0$) for $a \in \{\pm 0.4, \pm 0.6, \pm 0.8, \pm 1 \}$, $b=1$ (a) and for $a=1$, $b \in \{0.2, 0.4, 0.6, 0.8, 1 \}$ (b).} \label{fig:linex}
\end{figure}

Figures \ref{fig:linex} (a) and (b) show the LINEX function $L(\delta,0)$ for $a \in \{\pm 0.4, \pm 0.6, \pm 0.8, \pm 1 \}$, $b=1$ and for $a=1$, $b \in \{0.2, 0.4, 0.6, 0.8, 1 \}$ respectively. Higher absolute values of $a$ imply in higher losses. For example, in Figure \ref{fig:linex} (a), when $a=0.8$, $L(4,0) = 20.33$ while for $a = 1$, $L(4,0) = 49.59$. Thus the loss of an overestimation of 4 units under $a = 1$ is almost 2.44 times the loss under $a = 0.8$. Higher values of $b$ also increase the loss, as can be seen in Figure \ref{fig:linex} (b). For example, when $b = 0.8$, $L(4,0) = 39.68$, against $49.59$ when $b=1$. 

\subsection{Bayesian Shrinkage Rule }
\begin{figure}[H]
\centering
\subfigure[Bayesian rules.]{
\includegraphics[scale=0.55]{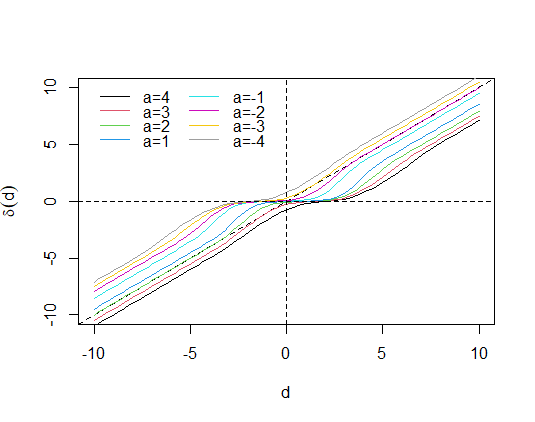}}
\subfigure[Bayesian shrinkage rules.]{
\includegraphics[scale=0.55]{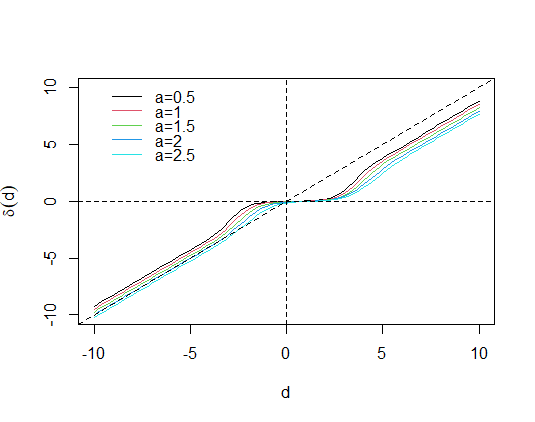}}
\caption{Bayesian rules \eqref{rule} for $\sigma = 1$, $\alpha = 0.9$, $\tau = 1$ and $a \in \{\pm 1, \pm 2, \pm 3, \pm 4 \}$ (a) and bayesian shrinkage rules (in wavelet shrinkage sense) for $a \in \{0.5, 1, 1.5, 2, 2.5 \}$ (b).} \label{fig:rules}
\end{figure}

The bayesian rule $\delta^{*}$ is obtained by minimizing the posterior expected loss $\mathbb{E}_{\pi}[L(\delta,\theta)|d]$. According to Zellner (1986), under the LINEX loss function we have
\begin{equation}
\mathbb{E}_{\pi}[L(\delta,\theta)|d] = b[e^{a\theta}\mathbb{E}_{\pi}(e^{-a\theta}|d) - a(\delta - \mathbb{E}_{\pi}(\theta|d)) - 1],
\end{equation}
and the bayesian rule is 
\begin{equation}\label{rule}
\delta^{*} = \delta^{*}(d) = (-1/a)\log(\mathbb{E}_{\pi}(e^{-a\theta}|d)),
\end{equation}
where $\log$ is the natural logarithm. Note that the bayesian rule under LINEX loss function does not depend on the parameter $b$. Finally, it is easy to show that under the model, one has that
\begin{equation}\label{exp}
\mathbb{E}_{\pi}(e^{-a\theta}|d) = \frac{\alpha \phi(d/\sigma) + \sigma (1-\alpha) \int_{\mathbb{R}} e^{-a(\sigma u + d)} g(\sigma u + d ; \tau) \phi(u) du}{\alpha \phi(d/\sigma) + \sigma (1-\alpha) \int_{\mathbb{R}} g(\sigma u + d ; \tau) \phi(u) du},
\end{equation}
where $\phi(\cdot)$ is the standard normal density function and the integrals involved in \eqref{exp} can be obtained numerically. Figure \ref{fig:rules} (a) shows the bayesian rules for $\sigma = 1$, $\alpha = 0.9$, $\tau = 1$ and $a \in \{\pm 1, \pm 2, \pm 3, \pm 4 \}$ on the interval $[-10,10]$. In this figure, it is possible to observe an important feature: the bayesian rule \eqref{rule} is not a shrinkage rule in the sense of wavelet shrinkage for all values of $a$, although the shape of the rules mimics wavelet shrinkage rules. For example, when $a = 4$, small values of the empirical coefficient $d$ are not shrunk toward zero by the bayesian rule. Further, if $d = 0$ then $\delta^{*}(0) = -0.768$. The main idea of a wavelet shrinkage rule is to shrink small empirical coefficients toward zero, even in an asymmetrically way, as it is the case under asymmetric loss functions or priors for the wavelet coefficients $\theta$. 

Hopefully, we can obtain shrinkage rules for convenient choices of $a$, as it is shown in Figure \ref{fig:rules} (b), for $a \in \{0.5, 1, 1.5, 2, 2.5\}$. Note that the rules shrink small empirical coefficients toward zero in an asymmetrically manner. Moreover, for empirical coefficients far from zero, the rule acts with different severity according to the sign of the coefficient. Since the LINEX loss penalizes exponentially the overestimation when $a > 0$, the associated shrinkage rules shrink severely empirical coefficients greater than zero to avoid overestimation of $\theta$, given the prior belief that most of the $\theta$'s are zero, and the severity increases for higher values of $a$. Although it is not shown in the paper, the behaviour of the rule is similar when $a < 0$, but in the opposite sense to avoid underestimation.  

\subsection{Statistical Properties of the Shrinkage Rule}

\begin{figure}[H]
\centering
\subfigure[Squared bias.]{
\includegraphics[scale=0.55]{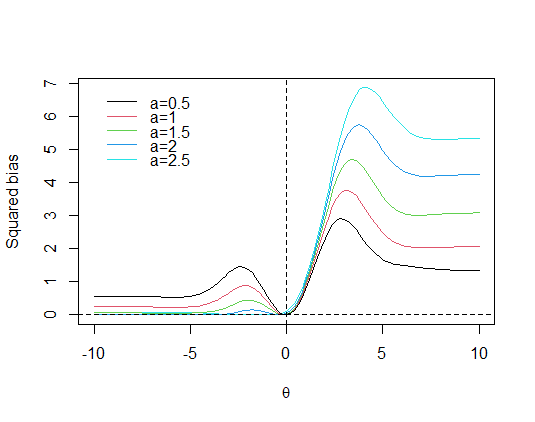}}
\subfigure[Variances.]{
\includegraphics[scale=0.55]{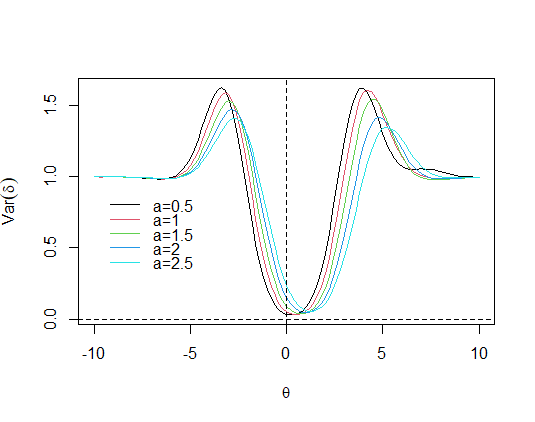}}
\subfigure[Frequentist risks.]{
\includegraphics[scale=0.55]{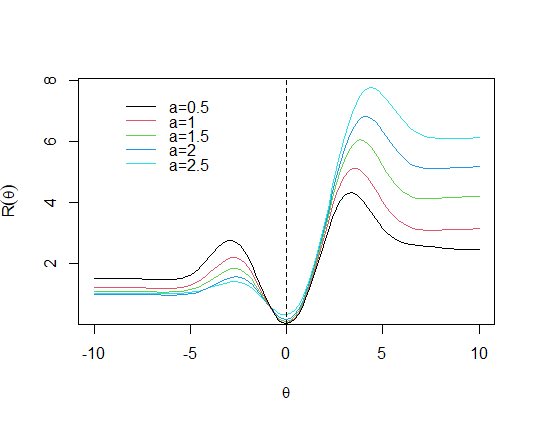}}
\caption{Squared bias (a), variances (b) and frequentist risks (c) of the shrinkage rules \eqref{rule} for $\sigma = 1$, $\alpha = 0.9$, $\tau = 1$ and $a \in \{0.5, 1, 1.5, 2, 2.5 \}$.} \label{fig:stats}
\end{figure}

The squared bias $\mathrm{Bias}^2_{\delta}(\theta) = [\mathbb{E}(\delta(d)) - \theta]^2$ of the shrinkage rules \eqref{rule} for $\sigma = 1$, $\alpha = 0.9$, $\tau = 1$ and $a \in \{0.5, 1, 1.5, 2, 2.5 \}$ were obtained numerically and are shown in Figure \ref{fig:stats} (a). In fact, the rules have small bias when $\theta$ is close to zero and the minimum squared bias is achieved when $\theta = 0$. Since LINEX loss is exponential for $\delta > \theta$ under positive values of $a$, i.e, the estimation procedure penalizes more overestimation, the squared bias is higher when $\theta$ is greater than zero. The shrinkage rule is built under the prior information that most of the wavelet coefficients are zero or close to zero. Then, for coefficients greater than zero, the shrinkage rule tends to underestimate them due exponential loss imposed by the LINEX loss function against overestimation. Further, the bias increases when $a$ increases, once the loss increases for higher $a$ values. On the other hand, the bias is small when $\theta$ is less than zero because the loss is only linear for $\delta < \theta$ and then, even under the prior belief of sparsity, the estimation procedure allows a good recover of wavelet coefficients less than zero. Analogous interpretation could be made for the frequentist risks $R_{\delta}(\theta) = \mathbb{E}[L(\delta,\theta)]$ of the rules \eqref{rule} and displayed in Figure \ref{fig:stats} (c).   

The variances $\mathrm{Var}_{\delta}(\theta) = \mathbb{E}[(\delta(d) - \mathbb{E}(\delta(d)))^2]$ of the shrinkage rules are provided in Figure \ref{fig:stats} (b). Unlike the squared bias and frequentist risks, the variances have symmetric shapes and are small when $\theta$ is close to zero. They increase as $|\theta|$ increases and achieve two maxima. Finally, the variances stabilize close to one for higher values of $|\theta|$. Further, the obtained maxima are higher as the LINEX parameter $a$ decreases. 

The general behaviours of the squared bias, variance and frequentist risk are quite similar. They are small when $\theta$ is close to zero, increase until two local maxima and then stabilize for sufficiently high values of $|\theta|$. Moreover, the two local maxima occur at the same locations where their associated shrinkage rules leave the x-axis and start to approximate to the line $y = x$. For example, when the LINEX parameter $a = 1$ (red curves in Figures \ref{fig:stats}), the global maxima of squared bias, variance and frequentist risk occur at $\theta = 3.12$ and are equals to 3.77, 1.12 and 4.99 respectively. Finally, as $\theta \rightarrow \infty$, $\mathrm{Bias}^2_{\delta}(\theta) \rightarrow 2.07$, $\mathrm{Var}_{\delta}(\theta)\rightarrow 1$ and $R_{\delta}(\theta) \rightarrow 3.09$. As $\theta \rightarrow -\infty$, $\mathrm{Bias}^2_{\delta}(\theta) \rightarrow 0.24$, $\mathrm{Var}_{\delta}(\theta)\rightarrow 1$ and $R_{\delta}(\theta) \rightarrow 1.22$.

\begin{table}[H]
\centering
\label{my-label}
\begin{tabular}{l|lllll}
\hline
$a$ & 0.5   & 1.0  &  1.5  & 2.0 & 2.5    \\ \hline 
$r_{\delta}$ & 0.1521 & 0.2005 & 0.2478 & 0.3512 & 0.4644  \\ \hline 
\end{tabular}
\caption{Bayes risks of the shrinkage rule \eqref{rule} for $\sigma = 1$, $\alpha = 0.9$, $\tau = 1$ and $a \in \{0.5, 1, 1.5, 2, 2.5 \}$.}\label{tab:brisk}
\end{table}

The Bayes risks $r_{\delta} = \mathbb{E}_{\pi}[R_{\delta}(\theta)]$ of the shrinkage rules are available in Table \ref{tab:brisk}. We note that the Bayes risk increases as the LINEX parameter $a$ increases. For example, for $a = 0.5$, we obtain $r_\delta = 0.1521$ and for $a = 2.5$ we have $r_\delta = 0.4644$, which is almost three times the risk for $a = 0.5$. Since shrinkage rules under higher values of $a$ shrink in a more severe way empirical coefficients greater than zero (once the loss is exponential), we have that the Bayes risks associated to more severe shrinkage rules are greater than less severe ones.  

\subsection{Comparison with Quadratic Loss Function}

The quadratic loss function $L(\delta,\theta) = (\delta - \theta)^2$ is perhaps the most considered loss function in statistical decision problems due to several reasons. Under a mathematical point of view, it is more tractably analytically than other loss functions and differentiable. Further, in statistical inference, when it is dealt with an unbiased estimator, the quadratic loss function becomes the variance of the estimator, with several nice properties and a rich theory, see Schervish (1995). 

\begin{figure}[H]
\centering
\subfigure[Loss functions.]{
\includegraphics[scale=0.43]{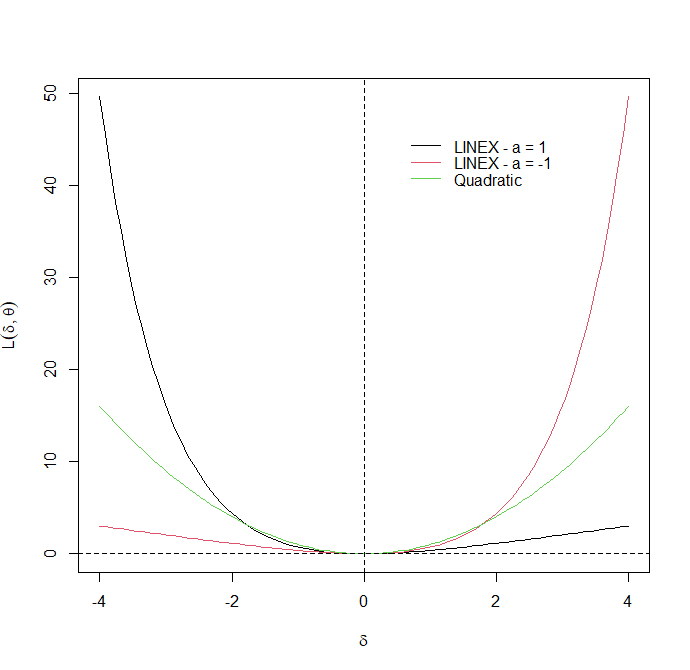}}
\subfigure[Shrinkage rules.]{
\includegraphics[scale=0.43]{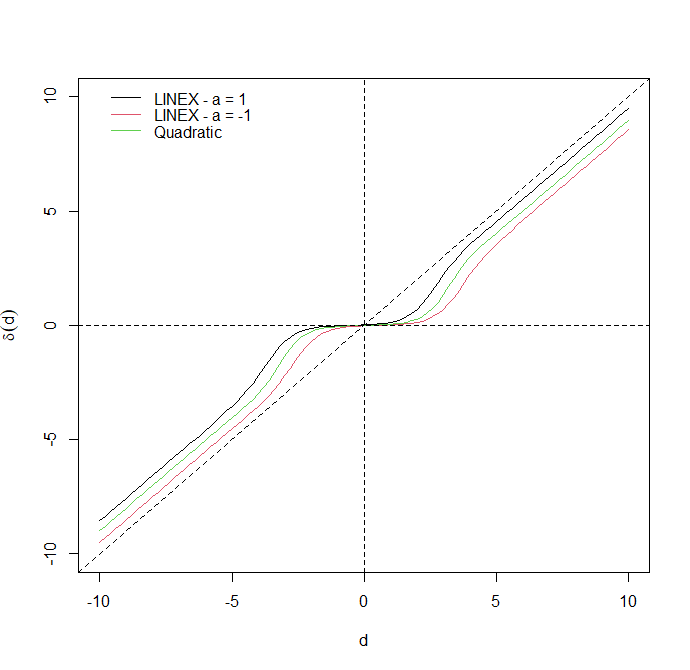}}
\caption{LINEX ($a = \pm 1$ and $b=1$) and quadratic loss functions for $\theta = 0$, i.e, $L(\delta,0)$ (a) and their associated shrinkage rules under the prior model \eqref{prior} for $\alpha = 0.9$ and $\tau = \sigma = 1$ (b).} \label{fig:comp}
\end{figure}

One important feature of the quadratic loss function is the symmetry around $\theta$, which means that underestimation and overestimation with the same magnitudes of a parameter $\theta$ by an estimator $\delta$ imply in the same polynomial (quadratic) loss. On the other hand, the LINEX loss function penalizes exponentially one side and linearly the other side depending on its parameters, i.e, the losses are assymmetric. 
Figure \ref{fig:comp}(a) presents the LINEX loss functions for $a = \pm 1$ and the quadratic loss function when $\theta = 0$. Under overestimation, i.e, when $\delta \gg 0$, the LINEX loss for $a = -1$ is considerably greater than the quadratic loss, once its loss increases exponentially. The LINEX loss for $a = 1$ increases only linearly and has the lowest loss. The same interpretation occurs for the underestimation case, but the roles of the LINEX rules change with each other.        

In the bayesian framework, the bayesian rule under the quadratic loss function is given by the posterior expected value of $\theta$, i.e, $\delta^{*}(d) = \mathbb{E}_{\pi}(\theta|d)$. In this case, Sousa (2020) showed that under the prior model \eqref{prior}, the shrinkage rule is given by
\begin{equation}
\delta(d) = \mathbb{E}_{\pi}(\theta|d) = \frac{(1-\alpha)\int_\mathbb{R}(\sigma u + d)g(\sigma u +d ; \tau)\phi(u)du}{\frac{\alpha}{\sigma}\phi(\frac{d}{\sigma})+(1-\alpha)\int_\mathbb{R}g(\sigma u +d ; \tau)\phi(u)du}. \nonumber
\end{equation} 

The shrinkage rules under LINEX loss for $a = \pm 1$ and under quadratic loss and prior model \eqref{prior} for $\alpha = 0.9$ and $\tau = \sigma = 1$ are shown in Figure \ref{fig:comp}(b). In fact, the shrinkage rule under quadratic loss acts symmetrically around zero and shrinks the empirical coefficients with intermediate severity in both sides. The shrinkage rule under LINEX loss with $a=-1$ shrinks empirical coefficients greater than zero with strong severity and the rule with $a=1$ is severe for coefficients less than zero. 

In general, the choice of the loss function has a great impact on the behaviour of the associated shrinkage rule. Symmetric loss functions like the quadratic assigns equals losses for underestimation and overestimation and imply in a symmetric shrinkage rule. Asymmetric loss functions like LINEX are suitable when the loss of underestimation is different to overestimation. In this case, the associated shrinkage rule is then asymmetric around zero.

\section{Simulation Studies}

Simulation studies were conducted to evaluate the performance of the shrinkage rule \eqref{rule} and to compare it against standard and recent proposed shrinkage and thresholding rules.  Wavelet coefficients were generated from a mixture of a point mass function at zero and the beta distribution with right asymmetries given by coefficient of kurtosis equals to 0.48 (scenario 1) and 1.73 (scenario 2) and two sample sizes, $n = 512$ and $2048$. Then, empirical wavelet coefficients were obtained from model \eqref{wmodel} with iid zero mean gaussian random errors and common variance $\sigma^2$ selected according to two signal to noise ratios (SNR), $\mathrm{SNR} = 3$ and $9$. For each replication $r$ , we applied the proposed shrinkage rule under LINEX loss and the comparison rules to the generated empirical coefficients and calculated the mean squared error (MSE) 
\begin{equation}
\mathrm{MSE}^{(r)} = \frac{1}{n} \sum_{i=1}^{n}(\hat{\theta}_i - \theta_i)^2, \nonumber
\end{equation}
and the mean absolute error (MAE)
\begin{equation}
\mathrm{MAE}^{(r)} = \frac{1}{n} \sum_{i=1}^{n}|\hat{\theta}_i - \theta_i|. \nonumber
\end{equation}
For each scenario of asymmetry, sample size and signal to noise ratio values, $200$ replications were run and the averaged mean squared error (AMSE) and the averaged mean absolute error (AMAE) were calculated as performance measures of the rules, given respectively by
\begin{equation}
\mathrm{AMSE} = \frac{1}{200} \sum_{r=1}^{200} \mathrm{MSE}^{(r)}, \nonumber
\end{equation}
and
\begin{equation}
\mathrm{AMAE} = \frac{1}{200} \sum_{r=1}^{200} \mathrm{MAE}^{(r)}. \nonumber
\end{equation}

We compared the proposed shrinkage rule (LINEX - Logistic) against the soft thresholding rules with the threshold value chosen according to the universal thresholding (Universal) proposed by Donoho and Johnstone (1994), cross validation (CV) by Nason (1996) and the Stein unbiased risk estimator (SURE) of Donoho and Johnstone (1995), the bayesian shrinkage rules under symmetric beta prior (Beta symmetric) proposed by Sousa et al. (2020) and asymmetric beta prior (Beta asymmetric) by Sousa (2021) and the soft thresholding rule under LINEX loss proposed by Torehzadeh and Arashi (2014).    

Table \ref{tab:sim1} shows the results of the scenario 1 under asymmetric distributed wavelet coefficients with kurtosis equals to 0.48. The proposed shrinkage rule had great performances for both values of sample sizes and SNR. Actually, the SNR had the main impact on the performances of the rules and the sample size did not have significant influence on them. For SNR = 3, the shrinkage rule under asymmetric beta prior was the best one in terms of AMSE and AMAE. The proposed rule and the shrinkage rule under symmetric beta prior also worked well and had performances close to the asymmetric beta rule one. SURE and the thresholding rule by Torehzadeh and Arashi had intermediate performances and Universal and CV thresholding rules did not worked well. For example, the AMSE and AMAE of the CV thresholding rule were respectively 10.24 and 3.38 times the AMSE and AMAE of the asymmetric beta shrinkage rule. For SNR = 9, the proposed rule was the best for both performance measures, with AMSE = 0.0004 and AMAE = 0.0121 when $n = 512$ and AMSE = 0.0003 and AMAE = 0.0113 when $n = 2048$. The shrinkage rule under asymmetric beta prior and SURE thresholding rule also had good behaviour, but with performances far for the proposed rule ones. For example, the SURE thresholding rule (the second better performance) had AMSE and AMAE respectively almost 3 and 2 times of the proposed rule ones when $n = 2048$. On the other hand, the thresholding rule by Torehzadeh and Arashi did not have good performance, which suggests that it works better for low signal to noise ratios.

\begin{table}[H]
\centering
\label{my-label}
\begin{tabular}{|c|c|c|c|}
\hline
 
 \textbf{n} & \textbf{Method} & \textbf{SNR=3}  & \textbf{SNR=9}   \\ \hline \hline
   &        & \textbf{AMSE (AMAE)} & \textbf{AMSE (AMAE)}   \\ \hline

512	&	Universal	&	0.0304	0.1141	&	0.0038	0.0416	\\
	&	CV	&	0.0420	0.1337	&	0.0056	0.0502	\\
	&	SURE	&	0.0090	0.0652	&	0.0010	0.0224	\\
	&	Beta symmetric	&	0.0045	0.0428	&	0.0049	0.0338	\\
	&	Torehzadeh and Arashi	&	0.0086	0.0629	&	1.8417	1.1194	\\
	&	LINEX - Logistic	&	0.0047	0.0428	&	\textbf{0.0004}	\textbf{0.0121}	\\
	&	Beta asymmetric	&	\textbf{0.0041}	\textbf{0.0396}	&	0.0007	0.0245	\\ \hline
									
2048	&	Universal	&	0.0337	0.1228	&	0.0043	0.0444	\\
	&	CV	&	0.0452	0.1423	&	0.0059	0.0525	\\
	&	SURE	&	0.0082	0.0620	&	0.0009	0.0210	\\
	&	Beta symmetric	&	0.0043	0.0426	&	0.0026	0.0436	\\
	&	Torehzadeh and Arashi	&	0.0076	0.0622	&	1.8517	1.1231	\\
	&	LINEX - Logistic	&	0.0045	0.0425	&	\textbf{0.0003}	\textbf{0.0113}	\\
	&	Beta asymmetric	&	\textbf{0.0039}	\textbf{0.0396}	&	0.0013	0.0318	\\ \hline

\end{tabular}
\caption{AMSE and AMAE of the shrinkage and thresholding rules for generated wavelet coefficients under asymmetric distribution with kurtosis equals to 0.48 - scenario 1.}\label{tab:sim1}
\end{table}

The results of the simulation study under scenario 2 is available in Table \ref{tab:sim2}. In fact, the proposed shrinkage rule had the best performance for the both measures in the four combinations of sample sizes and SNR. It suggests that the rule works better for denoising higher asymmetric distributed wavelet coefficients. The symmetric and asymmetric beta shrinkage rules also had good performances and so the SURE thresholding rule. It should be noted that the thresholding rule under LINEX loss by Torehzadeh and Arashi did not work well in these contexts. Further, its performance decreased for SNR = 9 under both sample size values. The standard thresholding rules under Universal and CV policies had intermediate performances relative to the comparison rules, but they were overcome by the proposed rule. For example, under $n = 512$ and SNR = 3, the AMSE and AMAE of CV rule were about 10.34 and 3.34 times of the proposed rule ones respectively. 

\begin{figure}[H]
\centering
\subfigure[MSEs - scenario 1.]{
\includegraphics[scale=0.4]{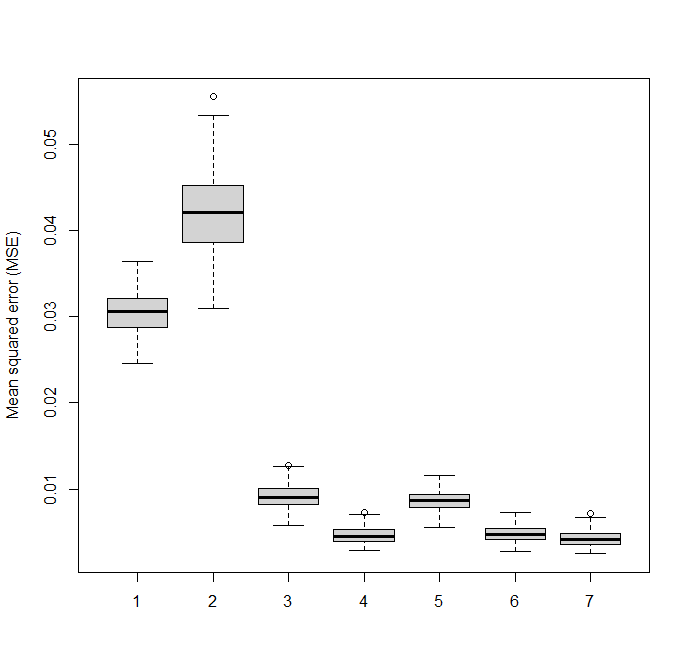}}
\subfigure[MAEs - scenario 1.]{
\includegraphics[scale=0.4]{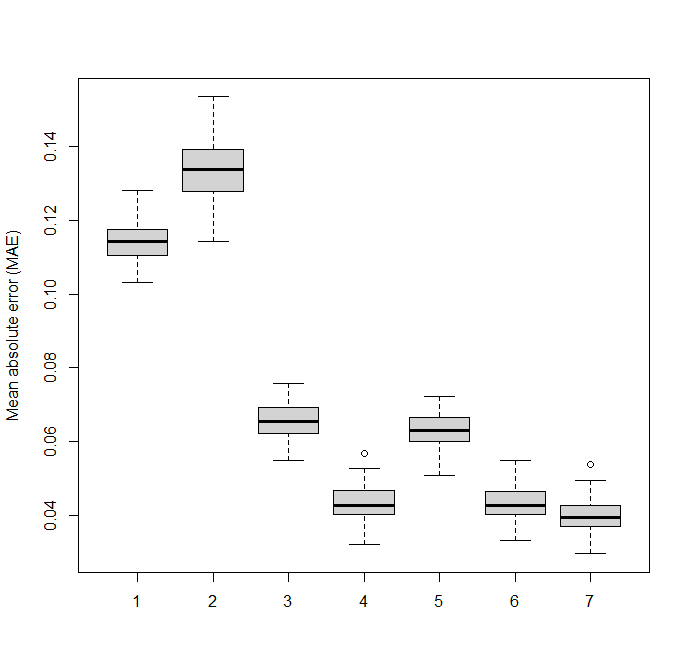}}
\subfigure[MSEs - scenario 2.]{
\includegraphics[scale=0.4]{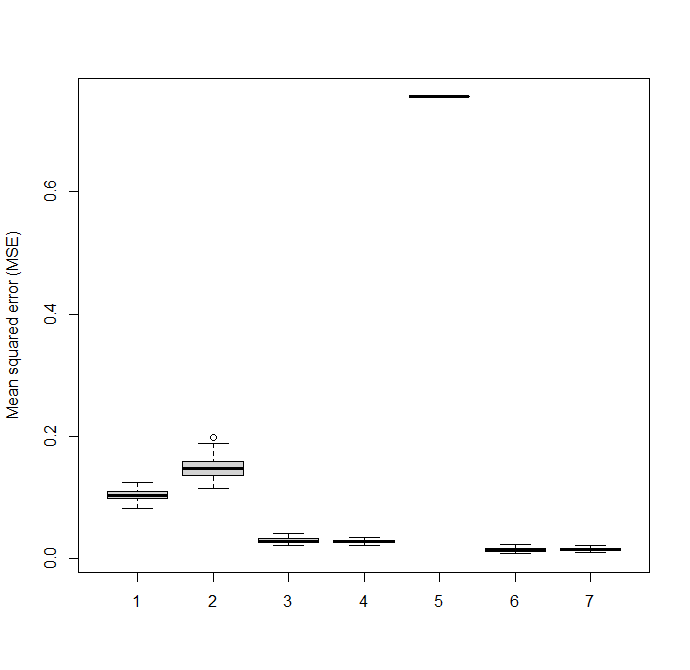}}
\subfigure[MAEs - scenario 2.]{
\includegraphics[scale=0.4]{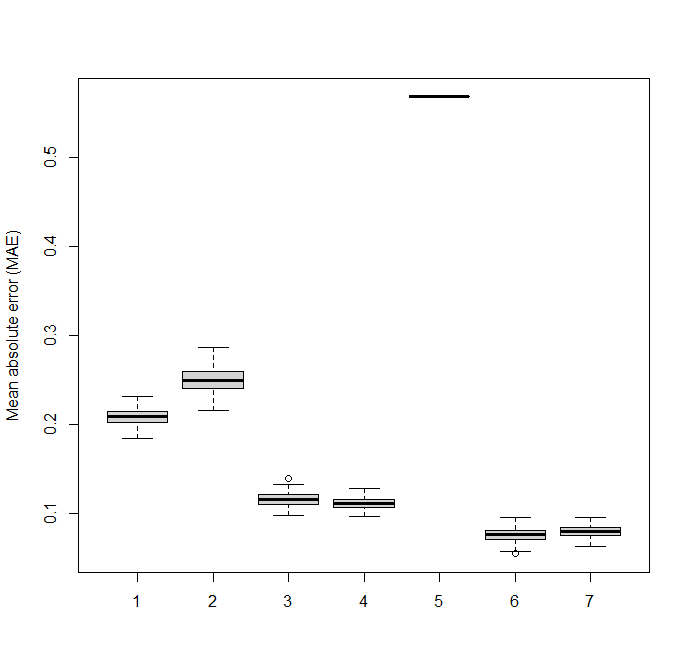}}
\caption{Boxplots of MSEs and MAEs of the shrinkage and thresholding rules of the simulation studies in scenario 1 (a) and (b) and scenario 2 (c) and (d) for $n=512$ and SNR = 3. The rules are: 1 - UNIV, 2 - CV, 3 - SURE, 4 - Beta symmetric, 5 - Torehzadeh and Arashi, 6 - LINEX - Logistic (proposed rule) and 7 - Beta asymmetric.} \label{fig:bp}
\end{figure}

Finally, Figure \ref{fig:bp} presents the boxplots of the MSEs and MAEs of the rules in the scenario 1 (a) and (b) and scenario 2 (c) and (d), for $n = 512$ and SNR = 3. The boxplots of scenario 1 show the well behaviour of the shrinkage rules under beta prior and the proposed rule in both measures. Universal and CV had high MSEs and MAEs relative to the others. On the other hand, the boxplots of scenario 2 show emphasizes the atypical behaviour of the thresholding rule of Torehzadeh and Arashi, with MSEs and MAEs considerably higher than the others ones. Similar behaviours were obtained for the other values of sample size and SNR. 

Thus, the simulation studies showed that the proposed shrinkage rule can be considered under asymmetric distributed wavelet coefficients context as an alternative to the shrinkage rule under asymmetric beta prior. Their performances were closed with each other and for higher kurtosis value, the proposed rule overcome the asymmetric beta shrinkage rule for both measures and all the combinations of sample size and SNR.

\begin{table}[H]
\centering
\label{my-label}
\begin{tabular}{|c|c|c|c|}
\hline
 
 \textbf{n} & \textbf{Method} & \textbf{SNR=3}  & \textbf{SNR=9}   \\ \hline \hline
   &        & \textbf{AMSE (AMAE)} & \textbf{AMSE (AMAE)}   \\ \hline

512	&	Universal	&	0.1021	0.2233	&	0.0113	0.0746	\\
	&	CV	&	0.1448	0.2668	&	0.0158	0.0885	\\
	&	SURE	&	0.0282	0.1189	&	0.0030	0.0394	\\
	&	Beta symmetric	&	0.0273	0.1191	&	0.0021	0.0319	\\
	&	Torehzadeh and Arashi	&	0.7657	0.6411	&	0.9632	0.8099	\\
	&	LINEX - Logistic	&	\textbf{0.0140}	\textbf{0.0799}	&	\textbf{0.0010}	\textbf{0.0208}	\\
	&	Beta asymmetric	&	0.0145	0.0837	&	0.0012	0.0236	\\ \hline
									
2048	&	Universal	&	0.1056	0.2117	&	0.0117	0.0707	\\
	&	CV	&	0.1451	0.2492	&	0.0160	0.0828	\\
	&	SURE	&	0.0237	0.1026	&	0.0026	0.0343	\\
	&	Beta symmetric	&	0.0237	0.1035	&	0.0019	0.0278	\\
	&	Torehzadeh and Arashi	&	0.7020	0.5552	&	1.0487	0.8451	\\
	&	LINEX - Logistic	&	\textbf{0.0115}	\textbf{0.0685}	&	\textbf{0.0008}	\textbf{0.0174}	\\
	&	Beta asymmetric	&	0.0126	0.0732	&	0.0010	0.0201	\\ \hline

\end{tabular}
\caption{AMSE and AMAE of the shrinkage and thresholding rules for generated wavelet coefficients under asymmetric distribution with kurtosis equals to 1.73 - scenario 2.}\label{tab:sim2}
\end{table}

\section{Application in Real Dataset}

\textit{Serenoa repens} or Saw Palmetto is a palm native from North America whose extracted essential oil is applied in several medical treatments, such as Benign Prostatic Hyperplasia (BPH) in men, disorder of urinary system, sexual impotence and others. When the essential oil is adulterated by the addition of substances such as olive oil, the contaminated oil losses its chemical and biological properties and becomes inactive against BPH for example. Chemists usually apply infrared (IR) spectroscopy on samples of Saw Palmetto essential oil to detect peaks of absorbance and classify a given sample in pure or adulterated. 

\begin{figure}[H]
\centering
\subfigure[Dataset.]{
\includegraphics[scale=0.4]{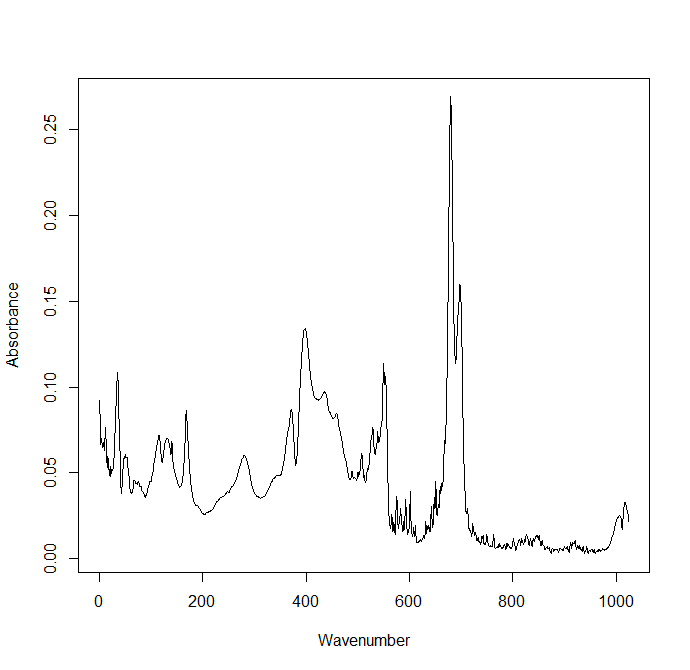}}
\subfigure[Empirical coefficients.]{
\includegraphics[scale=0.4]{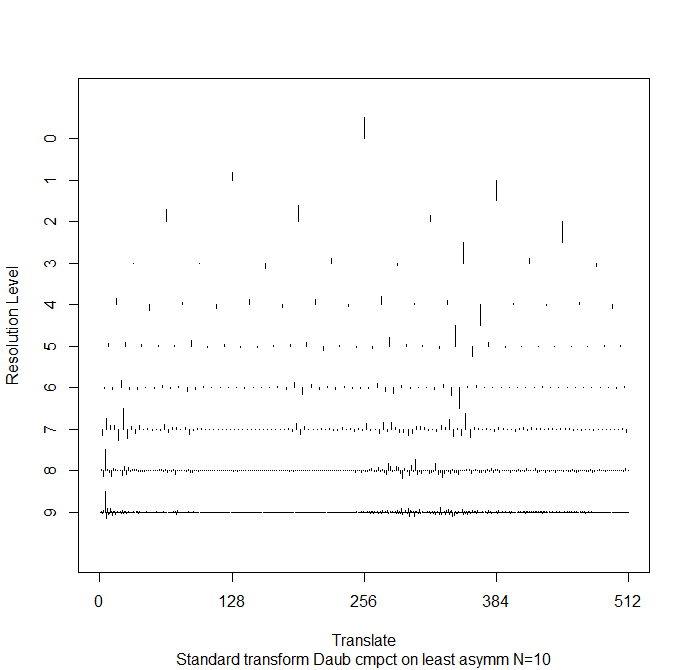}}
\caption{Infrared spectra of Saw Palmetto essential oil (a) and its empirical wavelet coefficients by resolution level after a DWT application (b).} \label{fig:app1}
\end{figure}

Figure \ref{fig:app1}(a) shows the IR spectra of a sample of pure Saw Palmetto oil in $n = 2^{10} = 1024$ points. This dataset is available in the R package \textit{ChemoSpec} by Hanson (2023). Actually the complete dataset consists of IR spectra of 14 samples of pure and adulterated samples in 1868 points. The particular data considered here is the IR spectra of the second oil sample and the considered sample size is the closest dyadic number that is smaller than 1868, i.e, $n = 1024 = 2^{\lfloor \log_2(1868) \rfloor}$. Since the goal is to obtain the peaks of absorbance, it is necessary to filter the data and remove noise that can lead to a misclassification of a peak. In this sense, the application of a wavelet shrinkage rule is suitable to denoise this data.

The first step is to obtain the DWT of the data. Figure \ref{fig:app1}(b) presents the magnitudes of the empirical wavelet coefficients of the dataset by resolution level. The considered wavelet basis was Daubechies with 10 null moments. In fact large empirical coefficients in the coarser resolution levels are associated to the positions of the absorbance peaks. On the other hand, the empirical coefficients in the finest resolution level are typically related to noise. The wavelet shrinkage rule acts mainly in these coefficients by reducing their magnitudes. 

\begin{figure}[H]
\centering
\subfigure[Denoised data.]{
\includegraphics[scale=0.4]{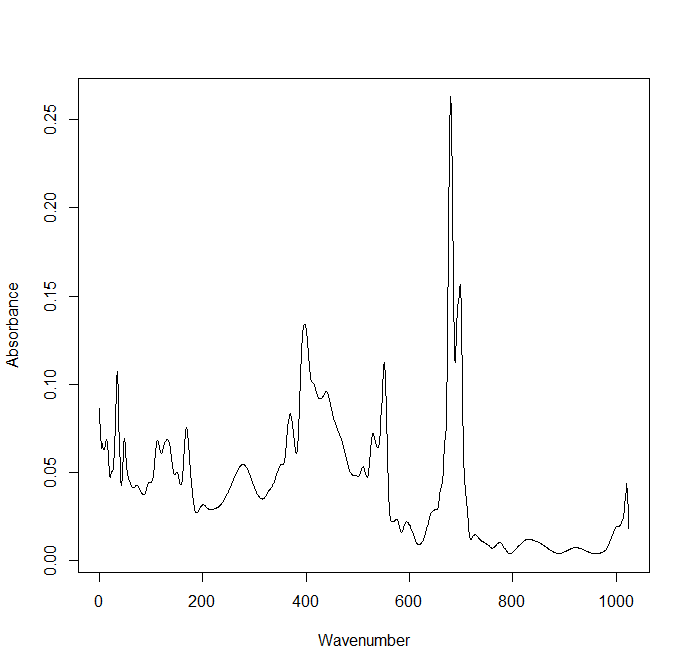}}
\subfigure[Empirical versus estimated coefficients.]{
\includegraphics[scale=0.4]{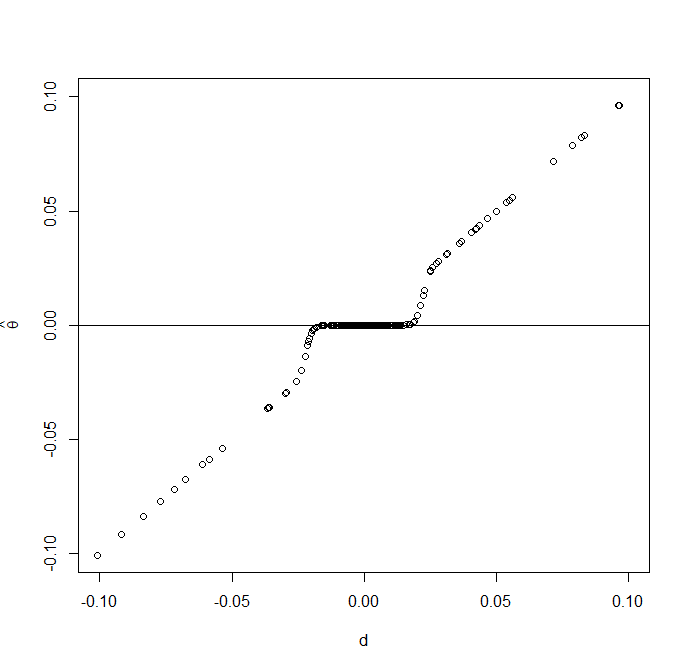}}
\caption{Denoised version of the infrared spectra of Saw Palmetto essential oil by the application of the shrinkage rule under LINEX loss function (a) and empirical coefficients in $[-0.10,0.10]$ versus estimated (shrunk) wavelet coefficients (b).} \label{fig:app2}
\end{figure}

The proposed shrinkage rule \eqref{rule} was applied in the empirical coefficients and the denoised version of the IR spectra was recover by the IDWT application \eqref{idwt}. The chosen parameter of the LINEX loss function was $a = 1$, the scale hyperparameter of the logistic density in the prior \eqref{prior} was $\tau = 5$, the hyperparameter $\alpha$ was chosen to be level dependent according to \eqref{eq:alpha} and the estimated standard deviation was $\hat{\sigma} = 0.0004$ according to \eqref{eq:sigma}. The obtained denoised IR spectra is showed in Figure \ref{fig:app2}(a). It is possible to observe the action of denoising in the wavenumber interval $[500,700]$ and $[800,1024]$, where local noise peaks were smoothed. Figure \ref{fig:app2}(b) presents a plot of the empirical coefficients in the interval $[-0.10,0.10]$ versus estimated wavelet coefficients. This plot is interesting to see the range of shrunk coefficients, which occurred mainly in the interval of empirical coefficients $[-0.05,0.05]$. Once the signal-to-noise ratio of the data is high, the shrinkage rule was active only in small empirical coefficients typically localized in the finer resolution levels. This procedure can be applied in the other samples for better identification of peaks and noise removal.

\section{Final Considerations}
The present work proposed a wavelet shrinkage rule under LINEX loss function and a mixture of a point mass function at zero and the logistic distribution as prior distribution to the wavelet coefficients. The application of this rule is suitable when overestimation and underestimation of the coefficients have asymmetric losses. In the wavelet domain, underestimated significant coefficients can lead to a bad detection of important features to be recovered of the unknown function such as peaks, discontinuities and oscillations. In this sense, to assign higher losses to underestimation should be interesting. Further, the simulation studies suggested a good performance of the proposed shrinkage rule in terms of averaged mean squared error and averaged mean absolute error measures to estimate asymmetrically distributed wavelet coefficients against standard shrinkage and thresholding rules. 

The magnitudes of the losses and the severity that the empirical coefficients are shrunk by the rule are well controlled according to convenient choices of the parameters of the LINEX function and the hyperparameters of the prior distribution, which is important to have flexibility in real datasets analysis. The impact of the wavelet basis in the estimation process and the proposal of other asymmetric loss functions are suggested as future works.

\end{document}